\documentclass[aps,nofootinbib,amsfonts,superscriptaddress,showpacs,showkeys]{revtex4}
\usepackage{amsfonts,amssymb,amscd,amsmath}
\usepackage{graphicx}
\DeclareMathOperator{\tr}{tr}
\DeclareMathOperator{\Tr}{Tr}

\newcommand{\mqr}{m^{\text q}_R}
\newcommand{\slashit}[1]{#1 \kern-.45em\slash}
\newcommand{\slashp}{\slashit p}
\newcommand{\slashP}{P \kern-.65em\slash }

\begin{document}
\title{ Baryon structure in a quark-confining non-local NJL model}
\author{Amir H. Rezaeian}
\email{Rezaeian@dirac.phy.umist.ac.uk}
\altaffiliation{Address after Nov. $1^{\text{st}}$, 2004: Institute for Theoretical Physics, University of Heidelberg,
Philosophenweg 19, D-69120 Heidelberg, Germany}
\affiliation{Department of Physics, UMIST, PO Box 88, Manchester, M60 1QD, UK}
\author{Niels R. Walet}
\email{Niels.walet@umist.ac.uk}
\affiliation{Department of Physics, UMIST, PO Box 88, Manchester, M60 1QD, UK}
\author{Michael C. Birse}
\email{Mike.birse@man.ac.uk}
\affiliation{Theoretical Physics Group, Department of Physics and Astronomy, University of Manchester, Manchester, M13 9PL, UK}
\date{\today}
\begin{abstract}
We study the nucleon and diquarks in a non-local Nambu-Jona-Lasinio
model. For certain parameters the model exhibits quark confinement, in
the form of a propagator without real poles. After truncation of the
two-body channels to the scalar and axial-vector diquarks,
a relativistic Faddeev equation for nucleon bound states is solved in
the covariant diquark-quark picture.  The dependence of the nucleon
mass on diquark masses is studied in
detail. We find parameters that lead to a simultaneous reasonable
description of pions and nucleons. Both the diquarks contribute attractively
to the nucleon mass. Axial-vector diquark correlations are seen to be
important, especially in the confining phase of the model. We study the possible implications
of quark confinement for the description of the diquarks and the
nucleon. In particular, we find that it leads to a more compact
nucleon. 
\end{abstract}
\maketitle
\date{\today}


\section{Introduction}
The NJL model is a successful phenomenological field theory inspired
by QCD \cite{njl}. The model is constructed to obey the basic
symmetries of QCD in the quark sector, but unlike the case of
low-energy QCD, quarks are not confined. The basic ingredient of the
model is a zero-range interaction containing four fermion fields.
This means that the model is not renormalizable. Therefore at one-loop
level an ultraviolet cut-off supplemented with a regularization method
is required from the outsets. The value of the cut-off can be related
to the scale of physical processes not included in the model, and thus
determines its range of validity.  Consequently, processes involving a
large momentum transfer can not be described by the model. At higher
orders in the loop expansion, which are necessary for calculating
mesonic (baryonic) fluctuations
\cite{njl-cut,cut-b}, one needs extra cut-off parameters. It is hard
to determine these parameters from independent physics, and thus to
build a viable phenomenology. A similar problem appears in the
diquark-quark picture of baryons where an additional cut-off parameter
is required to regularise the diquark-quark loops
\cite{cut-b}.

Another drawback of the model is the absence of confinement, which
makes it questionable for the description of few-quark states and for
quark matter.  If energetically allowed, the mesons of the model can
decay into free quark-antiquark pairs, and the presence of unphysical
channels is another limitation on the applicability of NJL model. At
the same time, it is also known that the NJL model exhibits a
zero-temperature phase transition at unrealistically low baryon
density \cite{njl-ma}. This problem is caused by the formation of
unphysical coloured diquark states. These may be explicitly excluded
at zero density by a projection onto the physical channels, but
dominate the behaviour at finite density.  The model is not able to
describe nuclear matter, even in the low-density regime \cite{w}.

We do not know how to implement colour confinement in the model and,
anyway, the exact confining mechanism of QCD is still unknown. In the
context of an effective quark theory, a slightly different mechanism
of ``quark confinement'' can be described by a quark propagator which
vanishes due to infra-red singularities \cite{rho2} or which does
not produce any poles corresponding to asymptotic quark states
\cite{rho1,asy}. Another realisation of quark confinement
can be found in Ref.~\cite{rho}. It has been shown that a non-local covariant extension of
the NJL model inspired by the instanton liquid model
\cite{non1} can lead to quark confinement for acceptable values of
the parameters \cite{pb}. This model has previously been applied to
mesons \cite{pb,pb-2,a1} and in this paper it is applied to baryons based on the relativistic Faddeev approach.

The quark propagator in the model has no real pole and consequently
quarks do not appear as asymptotic states. Instead the quark
propagator has pairs of complex poles. This phenomenon was also
noticed in Schwinger-Dyson equation studies in QED and QCD
\cite{sde-p,sde-new,sde-qe}. One can simply accept the appearance of
these poles as an artifact of the naive truncation scheme
involved. However, it has been recently suggested that it might be a
genuine feature of the full theory, and be connected with the
underlying confinement mechanism \cite{sde-new,sde-qe}. For example,
it has been shown by Maris that if one removes the confining potential
in QED in 2+1D the mass singularities are located almost on the time
axis, and if there is a confining potential, the mass-like
singularities move from the time axis to complex momenta
\cite{sde-qe}. In this paper, we study this kind of confinement from
another viewpoint. We show that when we have quark confinement in the
non-local NJL model, the baryons become more compact, compared to a
situation where we have only real poles for quark propagator.

There are several other advantages of the non-local version of the
model over the local NJL model: the dynamical quark mass is
momentum-dependent, as also found in lattice simulations of QCD
\cite{lat-m}. There various methods are available for construction of a conserved current in the presence of non-local interactions \cite{non-j}. 
In general, one can preserve the gauge invariance and anomalies by introducing 
additional non-local terms in the currents \cite{non-j}. A Noether-like method of
construction for these non-local pieces for the non-local NJL model was developed in Ref.~\cite{pb}.
The regulator makes the theory finite to all orders in the loop
expansion and leads to small next-to-leading order corrections
\cite{pb-2}.  As a result, the non-local version of the NJL model
should have more predictive power.

We use a separable non-local interactions, similar to that of the
instanton-liquid model \cite{non1,sep-i}.  This considerably simplifies the
calculation. Other approaches also give non-locality but in different forms \cite{asy,non-bir}. Non-locality also emerges naturally in the
Schwinger-Dyson resummation \cite{asy} and in various types of gluonic field configuration within the QCD vacuum, see for an example Ref.~\cite{gluon}.

Considerable work has been done on these nonlocal NJL models including
applications to the mesonic sector \cite{pb,pb-2,a1}, phase transitions
at finite temperature and densities \cite{a2}, and the study of chiral
solitons \cite{a3}.

In this paper we present our first results
from a calculation of the relativistic Faddeev equation for a
non-local NJL model, based on the covariant diquark-quark picture of baryons
\cite{njln1,njln2,njln3,njln4-is,njln4,o1,o2,o3,o4}. 
Such an approach has been extensively employed to study baryons in the
local NJL model, see, e.g.,
Refs.~\cite{njln1,njln2,njln3,njln4-is,njln4}.  We include both scalar
and the axial-vector diquark correlations. We do not assume a special
form for the interaction Lagrangian, but we rather treat the coupling
in the diquark channels as free parameters and consider the range of
coupling strengths which lead to a reasonable description of the
nucleon. We construct diquark and nucleon solutions and study the
possible implications of the quark confinement for the solutions. The
dependence of the baryon masses and waves on the diquarks parameters is investigated and
the role of diquarks in the nucleon solutions, for both the confining and
the non-confining phase of the model is considered separately.  The nucleon
wave function is studied in details.  Due to the separability of the
non-local interaction, the Faddeev equations can be reduced to a set
of effective Bethe-Salpeter equations. This makes it possible to adopt
the numerical method developed for such problems in
Refs.~\cite{o1,o2,o3,o4}.

This paper is organised as follows: In Sec.~II the model is
introduced. We also discuss the pionic sector of the model and
fix the parameters. In Sec.~III the diquark problem is solved and
discussed. In Sec.~IV the three-body problem based on diquark-quark
picture is investigated. The numerical technique involved in solving
the effective Bethe-Salpeter equation is given and the results for
three-body sector are presented. Finally, a summary and  outlook
is given in Sec.~V.

\section{A non-local NJL model\label{sec:model}}
We consider a non-local NJL model Lagrangian with $SU(2)_{f}\times
SU(3)_{c}$ symmetry.
\begin{equation}
\mathcal{L}=\bar{\psi}(i\slashit\partial-m_{c})\psi+\mathcal{L}_{I},
\end{equation}
where $m_{c}$ is the current quark mass of the $u$ and $d$ quarks
and $\mathcal{L}_{I}$ is a chirally invariant non-local
interaction Lagrangian. Here we restrict the interaction terms to
four-quark interaction vertices.

There exist several versions of such non-local NJL models. Regardless
of what version is chosen, by a Fierz transformation one can rewrite
the interaction in either the quark-antiquark or quark-quark channels. We
therefore use the interaction strengths in those channels as
independent parameters. For simplicity we truncate the mesonic
channels to the scalar ($0^{+},T=0$) and pseudoscalar ($0^{-},T=1$)
ones. The quark-quark interaction is truncated to the scalar ($0^{+},T=0$)
and axial vector ($1^{+},T=1$) colour $\overline{3}$ quark-quark channels (the
colour $6$ channels do not contribute to the colourless three-quark
state considered here). We parametrise the relevant part of
interaction Lagrangian as
\begin{eqnarray}\label{n1}
\mathcal{L}_{I}&=&\frac{1}{2}g_{\pi} j_{\alpha}(x)j_{\alpha}(x)+g_{s}\overline{J}_{s}(x)J_{s}(x)+g_{a}\overline{J}_{a}(x)J_{a}(x),\nonumber\\
j_{\alpha}(x)&=&\int
d^{4}x_{1}d^{4}x_{3}F(x-x_{3})F(x_{1}-x)\overline{\psi}(x_{1})\Gamma_{\alpha}\psi(x_{3}),\nonumber\\
\overline{J}_{s}(x)&=&\int
d^{4}x_{1}d^{4}x_{3}F(x-x_{3})F(x_{1}-x)\overline{\psi}(x_{1})\big[\gamma_{5}C\tau_{2}\beta^{A}\big]\overline{\psi}^{T}(x_{3}),\nonumber\\
J_{s}(x)&=&\int
d^{4}x_{2}d^{4}x_{4}F(x-x_{4})F(x_{2}-x)\psi^{T}(x_{2})\big[C^{-1}\gamma_{5}\tau_{2}\beta^{A}\big]\psi(x_{4}),\nonumber\\
\overline{J}_{a}(x)&=&\int
d^{4}x_{1}d^{4}x_{3}F(x-x_{3})F(x_{1}-x)\overline{\psi}(x_{1})\big[\gamma_{\mu}C\tau_{i}\tau_{2}\beta^{A}\big]\overline{\psi}^{T}(x_{3}),\nonumber\\
J_{a}(x)&=&\int
d^{4}x_{2}d^{4}x_{4}F(x-x_{4})F(x_{2}-x)\psi^{T}(x_{2})\big[C^{-1}\gamma^{\mu}\tau_{2}\tau_{i}\beta^{A}\big]\psi(x_{4}),\
\end{eqnarray}
where $\Gamma_{\alpha}=(1,i\gamma_{5}\tau)$. The matrices
$\beta^{A}=\sqrt{3/2} \lambda^{A}(A=2, 5, 7)$ project onto the colour
$\overline{3}$ channel with normalisation
$\tr(\beta^{A}\beta^{A'})=3\delta^{AA'}$ and the ${\tau_{i}}$'s are
flavour $SU(2)$ matrices with $\tr
(\tau_{i}\tau_{j})=2\delta_{ij}$. The object $C=i\gamma_{2}\gamma_{5}$
is the charge conjugation matrix.  It is exactly this four-way separability of the non-local interaction that is also
present in the instanton liquid model \cite{sep-i}.

Since we do not restrict ourselves to a specific choice of underlying interaction,
we shall treat the couplings $g_{s}$, $g_{a}$ and $g_{\pi}$ as
independent parameters. We assume $g_{\pi,s,a}>0$, which leads to
attraction in the given channels (and repulsion in
the quark-antiquark colour octet and quark-quark colour antisextet channels). The
coupling parameter $g_{\pi}$ is responsible for the pions and their
isoscalar partner $\sigma$. The coupling strengths $g_{s}$ and $g_{a}$
specify the behaviour in the scalar and axial-vector diquark channel,
respectively.

We define the Fourier transform of the form factor by 
\begin{equation}
F(x-x_{i})=\int \frac{d^{4}p}{(2\pi)^{4}} e^{-i(x-x_{i})\cdot p}f(p). \label{forfa}
\end{equation}
The dressed quark
propagator $S(k)$ is now constructed by means of a Schwinger-Dyson
equation (SDE) in the rainbow-ladder approximation. Thus the
dynamical constituent quark mass, arising from spontaneously broken
chiral symmetry, is obtained  in Hartree approximation as\footnote{The symbol
$\Tr$ denotes a trace over flavour, colour and Dirac indices and
$\tr_{D}$ denotes a trace over Dirac indices only.}
\begin{equation}
m(p)=m_{c}+ig_{\pi}f^{2}(p)\int \frac{d^{4}k}{(2\pi)^{4}} \Tr 
[S(k)] f^{2}(k), \label{n3-gap}
\end{equation}
where
\begin{equation}
S^{-1}(k)=\slashit{k}-m(k). \label{n3}
\end{equation}
One can simplify this equation by writing $m(p)$ in the form
\begin{equation}
m(p)=m_{c}+(m(0)-m_{c})f^{2}(p).  \label{nI}
\end{equation}
The non-linear equation can then be solved iteratively for $m(0)$.

Following Ref.~\cite{pb}, we choose the form factor to be Gaussian in
Euclidean space, $f(p_{E})=\exp(-p_{E}^{2}/\Lambda^{2})$, where
$p_{E}$ denotes the Euclidean four-momentum and $\Lambda$ is a cutoff
of the theory.  This choice respects Poincar\'e invariance and for
certain values of the parameters it leads to quark, but not colour,
confinement.  For values of $m(0)$ satisfying
\begin{equation} \label{n4}
\frac{m(0)-m_{c}}{\sqrt{m^{2}_{c}+\Lambda^{2}}-m_{c}} >
\frac{1}{2}\exp\left(-\frac{(\sqrt{m^{2}_{c}+\Lambda^{2}}+m_{c})^{2}}{2\Lambda^{2}}\right),
\end{equation}
the dressed quark propagator has no poles at real $p^{2}$ in Minkowski
space ($p^{2}+m^{2}(p^{2})\neq 0$).  The propagator has many pairs of
complex poles, both for confining and non-confining parameter
sets. This is a feature of these models and due care should be taken
in handling such poles, which can not be associated with asymptotic
states if the theory is to satisfy unitarity.  One should note that
the positions of these poles depend on the details of the chosen form
factor and the cut-off, hence one may regard them as a pathology of
the regularisation scheme.  Since the choice of the cut-off is closely
related to the truncation of the mesonic channels, (for example, if
one allows mixing of channels, the cut-off and the positions of poles
will change).  Even though the confinement in this model has no direct
connection to the special properties of the pion, there is an indirect
connection through the determination of the parameters from the pionic
properties.
\begin{figure}
\centerline{\includegraphics[clip,width=10 cm]{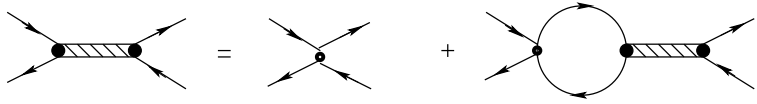}}
\caption{A graphical representation of the Bethe-Salpeter equation for the 
$\bar{q}q$ $T$-matrix in RPA approximation. The solid lines denote the
dressed quark propagators Eq.~(\ref{n3}) and shaded boxes denote meson
propagators.\label{fig:meson}}
\end{figure}
\begin{figure}
\centerline{\includegraphics[clip,width=10 cm]{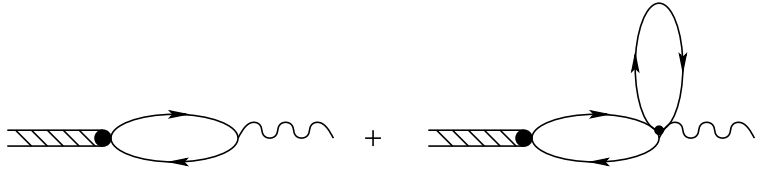}}
\caption{One-pion-to-vacuum matrix element in RPA, contributing to the weak pion decay. The lines are as defined in Fig.~\ref{fig:meson}. The wavy line denotes a weak decay.\label{fig:fpi}}
\end{figure}

The quark-antiquark $T$-matrix in the pseudoscalar channel can be
solved by using the Bethe-Salpeter equation in the random phase
approximation (RPA), as shown in Fig.~\ref{fig:meson}, see
Ref.~\cite{pb}.
\begin{eqnarray}
T(p_{1},p_{2},p_{3},p_{4})&=&f(p_{1})f(p_{2})\big[i\gamma_{5}\tau_{i}\big]\frac{g_{\pi}i}{1+g_{\pi}J_{\pi}(q^{2})}\big[i\gamma_{5}\tau_{i}\big]
f(p_{3})f(p_{4})\label{pi1}\nonumber\\
&&\times\delta(p_{1}+p_{2}-p_{3}-p_{4}),\
\end{eqnarray}
where 
\begin{eqnarray}
 J_{\pi}(q^{2})&=& i \Tr \int
 \frac{d^{4}k}{(2\pi)^{4}}f^{2}(k)\gamma_{5}\tau_{i}S(k)\gamma_{5}\tau_{i}S(q+k)f^{2}(q+k),\nonumber\\
&=&6i\int\frac{d^{4}k}{(2\pi)^{4}}\tr_{D}[\gamma_{5}S(k)\gamma_{5}S(k+q)]f^{2}(k)f^{2}(q+k),\label{pi2}\
\end{eqnarray}
where $q$ denotes the total momentum of the quark-antiquark pair. The pion
mass $m_{\pi}$ corresponds to the pole of $T$-matrix. One 
immediately finds that $m_{\pi}=0$ if the current quark mass $m_{c}$
is zero, in accordance with Goldstone's theorem. The residue of the
$T$-matrix at this pole has the form
\begin{equation}
V^{\pi}(p_{1}, p_{2})=ig_{\pi qq}[\openone_{c}\otimes \tau^{a}\otimes\gamma_{5}]f(p_{1})f(p_{2}),    \label{pi3}      
\end{equation}
where $g_{\pi qq}$ is the pion-quark-antiquark coupling
constant and is related to the corresponding loop integral $J_\pi$ by
\begin{equation}
g^{-2}_{\pi qq}=\left.\frac{dJ_{\pi}}{dq^{2}}\right|_{q^{2}=m^{2}_{\pi}}.
\end{equation}

\begin{table}
\caption{The parameters for the sets $A$ and $B$,
fitted to $f_{\pi}=92.4$ MeV and $m_{\pi}=139.6$ MeV. The resulting values of the
dynamical quark mass $m(0)$ are also shown.\label{tab:modpar}}
\begin{ruledtabular}
\begin{tabular}{lll}
Parameter &set A & set B  \\
\hline
$m(0)$ (MeV) & 297.9& 351.6  \\
$m_{0}(0)$ (MeV) & 250 & 300 \\
$m_{c}$ (MeV) & 7.9 & 11.13 \\
$\Lambda$ (MeV) & 1046.8 &847.8 \\
$g_{\pi} (\text{GeV}^{-2})$ &31.6 & 55.80 \\
\end{tabular}
\end{ruledtabular}
\end{table}

\begin{table}
\caption{The first two sets of  poles of the quark propagator (in  magnitude) in
the Minkowski frame. \label{tab:quarkpoles}}
\centering
\begin{ruledtabular}
\begin{tabular}{ll}
set A  & set B \\
\hline
$\pm 391$ MeV&$\pm 408 \pm 238i$ MeV\\
$\pm 675$ MeV& $\pm 1575\pm 307i$ MeV\\
\end{tabular}
\end{ruledtabular}
\end{table}

The pion decay constant $f_{\pi}$ is obtained from the coupling of the
pion to the axial-vector current. Notice that due to the non-locality
the axial-vector current is modified \cite{pb,non-j} and consequently
the one-pion-to-vacuum matrix element gets the additional contribution
shown in Fig.~\ref{fig:fpi}. This extra term is essential in order to
maintain Gell-Mann-Oakes-Renner relation \cite{pb} and makes a
significant contribution. The pion decay constant is
given by
\begin{eqnarray}
f_{\pi}&=&\frac{ig_{\pi \bar{q}q}}{m_{\pi}^{2}}\int
\frac{d^{4}k}{(2\pi)^{4}}\Tr[\slashit{q}\gamma_{5}\frac{\tau_{a}}{2}(S(p_{-}))\gamma_{5}\tau_{a}(S(p_{+}))]f(p_{-})f(p_{+})\nonumber\\
&+&\frac{ig_{\pi}}{2m^{2}_{\pi}}\int
\frac{d^{4}k}{(2\pi)^{4}}\Tr[S(k)]\int
\frac{d^{4}k}{(2\pi)^{4}}\Tr[V^{\pi}(p_{-},p_{+})S(p_{-})\gamma_{5}\tau_{a}S(p_{+})]\nonumber\\
&&\times
[f^{2}(k)\left(f^{2}(p_{+})+f^{2}(p_{-})\right)-f(p_{+})f(p_{-})f(k)\left(f(k+q)+f(k-q)\right)], \label{fpi}\
\end{eqnarray}
where $V_{\pi}(p_{-},p_{+})$ is defined in Eq.~(\ref{pi3}), with the notation $p_{\pm}=p\pm\frac{1}{2}q$.

The loop integrations in Eqs.~(\ref{pi2},\ref{fpi}) are evaluated in
Euclidean space\footnote{We work in Euclidean space with metric
$g^{\mu\nu}=\delta^{\mu\nu}$ and a hermitian basis of Dirac matrices
$\{\gamma_{\mu},\gamma_{\nu}\}=2\delta_{\mu\nu}$, with a standard
transcription rules from Minkowski to Euclidean momentum space:
$k^{0}\to ik_{4} $, $\vec{k}^{M}\to -\vec{k}^{E}$}. For the current
model, the usual analytic continuation of amplitudes from Euclidean to
Minkowski space can not be used. This is due to the fact that quark
propagators of the model contain many poles at complex energies
leading to opening of a threshold for decay of a meson 
into other unphysical states.  Any theory of this type needs an alternative continuation prescription consistent with
unitarity and macrocausality. Let us define a fictitious two-body
threshold as twice the real part of the first pole of the dressed quark propagator $\mqr$. For a confining parameter set, each quark
propagator has a pair of complex-conjugate poles. Above the two-body
pseudo-threshold $q^{2}<-4(\mqr)^{2}$, where $q$ is the meson 
momentum, the first pair of complex poles of the quark propagator has
a chance to cross the real axis. According to the Cutkosky
prescription \cite{cutk}, if one is to preserve the unitarity and the
microcausality, the integration contour should be pinched at that
point. In this way, one can ensure that there is no spurious
quark-antiquark production threshold, for energies below the next
pseudo-threshold, i.e. twice the real part of the second pole of the
quark propagator. Note that it has been shown \cite{u} that the
removal of the quark-antiquark  pseudo-threshold is closely related to the
existence of complex poles in the form of complex-conjugate pairs.
Since there is no unique analytical continuation method available for
such problems, any method must be regarded as a part of the model
assumptions \cite{pb,a1,u}. Here, we follow the method used in Ref.~\cite{pb}.

Our model contains five parameters: the current quark mass $m_{c}$,
the cutoff ($\Lambda$), the coupling constants $g_{\pi}$ , $g_{s}$ and
$g_{a}$. We fix the first three to give a pion mass of $m_{\pi}=139.6$
MeV with decay constant $f_{\pi}=92.4$ MeV, while we take the value of
the zero-momentum quark mass in the chiral limit $m_{0}(0)$ as an
input. We analyse two sets of parameters, as indicated in Table
\ref{tab:modpar}, where set $A$ is a non-confining parameter set,
while set $B$ leads to quark confinement (i.e., it satisfies the
condition Eq.~(\ref{n4})). The position of the quark poles are given
in Table \ref{tab:quarkpoles}. The real part of the first pole of
the dressed quark propagator $\mqr$ can be considered in much the same as the
quark mass in the ordinary NJL model. Since we do not believe in
on-shell quarks or quark resonances, this is also a measure for a
limit on the validity of the theory.  The mass $\mqr$ is larger than the constituent quark mass at zero momentum
$m(0)$, as can be seen in Table
\ref{tab:modpar}. As we will see $\mqr$ appears as an important parameter in the diquark and
nucleon solution, rather than the constituent quark mass. The same
feature has been seen in the studies of the soliton in this model, where
$\mqr$ determines the stability of the soliton
\cite{a3}. The parameters $g_{s}$ and $g_{a}$ will be treated here as free parameters, which allows us to
analyse baryon solutions in terms of a complete set of couplings. This
is permissible as long as the interactions in Lagrangian are not fixed
by some underlying theory via a Fierz transformation\footnote{Notice as well that the
Hartree-Fock approximation is equivalent to the Hartree approximation
with properly redefining coupling constants. Therefore, the Hartree
approximation here is as good as the Hartree-Fock one, since the interaction terms are not fixed by a Fierz transformation.}. 
The coupling-constant dependence is expressed through the ratios
$r_{s}=g_{s}/g_{\pi}$ and $r_{a}=g_{a}/g_{\pi}$. 

The quark condensate $\langle \bar{\psi}\psi\rangle=i\Tr S(0)$ is
closely related to the gap equation, Eq.~(\ref{n3-gap}). In the latter
there appears an extra form factor inside the loop integral. The quark
condensate in the chiral limit is $-(207 \text{ MeV})^{3}$ and
$-(186\text{ MeV})^{3}$ for sets A and B, respectively.
These values fall within the limits extracted from QCD sum rules
\cite{qcdsum} and lattice calculations \cite{lat-con}, having in mind
that QCD condensate is a renormalised and scale-dependent quantity.
In contrast to the local NJL model, here the dynamical quark mass
Eq.~(\ref{nI}) is momentum dependent and follows a trend similar to
that estimated from lattice simulations \cite{lat-m}.

\section{Diquark channels\label{sec:diquark}}
In the rainbow-ladder approximation the scalar quark-quark $T$-matrix can be
calculated from a very similar diagram to that  shown in
Fig.~\ref{fig:meson} (the only change is that the anti-quark must be
replaced by a quark with opposite momentum). It can be written as
\begin{eqnarray}
T(p_{1},p_{2},p_{3},p_{4})&=&f(p_{1})f(p_{2})\big[\gamma_{5}C\tau_{2}\beta^{A}\big]\tau(q)\big[C^{-1}\gamma_{5}\tau_{2}\beta^{A}\big]f(p_{3})f(p_{4})\label{tm1}\nonumber\\
&&\times\delta(p_{1}+p_{2}-p_{3}-p_{4}),\
\end{eqnarray}
with
\begin{equation}
\tau(q)=\frac{2g_{s}i}{1+g_{s}J_{s}(q^{2})},\label{n5-0}
\end{equation}
where $q=p_{1}+p_{2}=p_{3}+p_{4}$ is the total momentum of the quark-quark
pair, and
\begin{eqnarray}
 J_{s}(q^{2})&=& i \Tr \int
 \frac{d^{4}k}{(2\pi)^{4}}f^{2}(-k)\big[\gamma_{5}C\tau_{2}\beta^{A}\big]S(-k)^{T}\big[C^{-1}\gamma_{5}\tau_{2}\beta^{A}\big]S(q+k)f^{2}(q+k),\nonumber\\
&=&6i\int\frac{d^{4}k}{(2\pi)^{4}}\tr_{D}[\gamma_{5}S(k)\gamma_{5}S(k+q)]f^{2}(k)f^{2}(q+k)\label{l1}.
\end{eqnarray}
In the above equation the quark propagator $S(k)$ is the solution of
the rainbow SDE Eq.~(\ref{n3}).  The denominator of Eq.~(\ref{n5-0})
is the same as in the expression for the pion channel,
Eq.~(\ref{pi1}), if $g_{s}=g_{\pi}$. One may thus conclude that at
$r_{s}=1$ the diquark and pion are degenerate. This puts an upper
limit to the choice of $r_{s}$, since diquarks should not condense in
vacuum. One can approximate $\tau(q)$ by an effective diquark
``exchange'' between the external quarks, and parametrise $\tau(q)$ near
the pole as
\begin{equation}
\tau(q)=2ig^{2}_{dsqq}V^{s}(q)D(q), \hspace{2cm} D^{-1}(q)=q^{2}-M^{2}_{ds},\label{n5}
\end{equation}
where $M_{ds}$ is the scalar diquark mass, defined as the position of
the pole of $\tau(q)$.
The
strength of the on-shell coupling of scalar diquark to quarks,
$g_{dsqq}$ is related to the polarisation $J_{s}$ by
\begin{equation}
g^{-2}_{dsqq}=\frac{dJ_{s}}{dq^{2}}|_{q^{2}=M^{2}_{ds}},\label{co}
\end{equation}
and $V^{s}(q)$ is the ratio between the exact $T$-matrix and on-shell
(one-pole) approximation and describes the off-shell correction of the $T$-matrix
around the diquark solutions \cite{off-shell-d}. For the ``on-shell approximation''
we have $V^{s}(q)=1$ \cite{njl,cut-b}, and by definition on the mass shell $q^{2}=M^{2}_{ds}$, one has 
$V^{s}(q)|_{q^{2}=M^{2}_{ds}}=1$.

\begin{figure}
\centerline{\includegraphics[clip,width=7 cm]{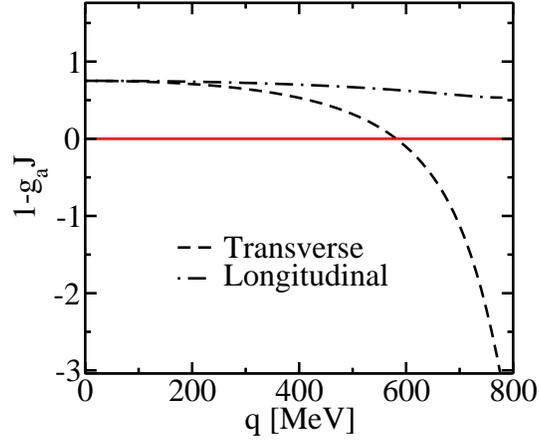}}
\caption{The denominator of the diquark $T$ matrix for the longitudinal and 
transverse axial vector channel. for parameter set A at $r_{a}=0.44$.
Note that there is no longitudinal pole.
\label{fig:avT}}
\end{figure}

There is no mixing between the axial-vector diquark and 
other channels, and so one can write the axial-vector
diquark $T$-matrix in a similar form
\begin{eqnarray}
T(p_{1},p_{2},p_{3},p_{4})&=&f(p_{1})f(p_{2})\big[\gamma_{\mu}C\tau_{i}\tau_{2}\beta^{A}\big]\tau^{\mu\nu}(q)\big[C^{-1}\gamma_{\nu}\tau_{2}\tau_{i}\beta^{A}\big]f(p_{3})f(p_{4})\label{tm2}\nonumber\\
&&\times\delta(p_{1}+p_{2}-p_{3}-p_{4}),\
\end{eqnarray}
with
\begin{equation}
\tau^{\mu\nu}(q)=2g_{a}i\Big[\frac{g^{\mu\nu}-q^{\mu}q^{\nu}/q^{2}}{1+g_{a}J^{T}_{a}(q^{2})}+\frac{q^{\mu}q^{\nu}/q^{2}}{1+g_{a}J^{L}_{a}(q^{2})}\Big],
\end{equation}
where we decompose the axial polarisation tensor into
longitudinal and transverse components:
\begin{eqnarray}
J^{\mu\nu}_{a}(q^{2})&=& i \Tr \int
\frac{d^{4}k}{(2\pi)^{4}}f^{2}(-k)\big[\gamma^{\mu}C\tau_{i}\tau_{2}\beta^{A}\big]S(-k)^{T}\big[C^{-1}\gamma^{\nu}\tau_{2}\tau_{i}\beta^{A}\big]S(q+k)f^{2}(q+k),\nonumber\\
&=&6i\int\frac{d^{4}k}{(2\pi)^{4}}\tr_{D}[\gamma^{\mu}S(k)\gamma^{\nu}S(k+q)]f^{2}(k)f^{2}(q+k)\nonumber\\
&=&J^{T}_{a}(q^{2})(g^{\mu\nu}-q^{\mu}q^{\nu}/q^{2})+J^{L}_{a}(q^{2})q^{\mu}q^{\nu}/q^{2}.\label{lon}\
\end{eqnarray}
We find  that the longitudinal channel does not
produce a pole (see Fig.~\ref{fig:avT}),
and thus  the bound axial-vector diquark solution corresponds to a pole of the
transverse $T$-matrix. The transverse component of 
$\tau^{\mu\nu}(q)$ matrix can be parametrized as,
\begin{equation} 
\tau^{\mu\nu}(q)= 2ig^{2}_{daqq}V^{a}(q)D^{\mu\nu}(q), \hspace{2cm} D^{\mu\nu}(q)=\frac{g^{\mu\nu}-q^{\mu}q^{\nu}/q^{2}}{q^{2}-M^{2}_{da}},\label{aloop}
\end{equation}
where $V^{a}(q)$ includes the off-shell contribution to the
axial-vector $T$-matrix. The coupling constant $g_{daqq}$ is related to the
residue at the pole of the $T$-matrix,
\begin{equation}
g^{-2}_{daqq}=\frac{dJ^{T}_{a}}{dq^{2}}|_{q^{2}=M^{2}_{da}}.\label{coa}
\end{equation}
\begin{figure}
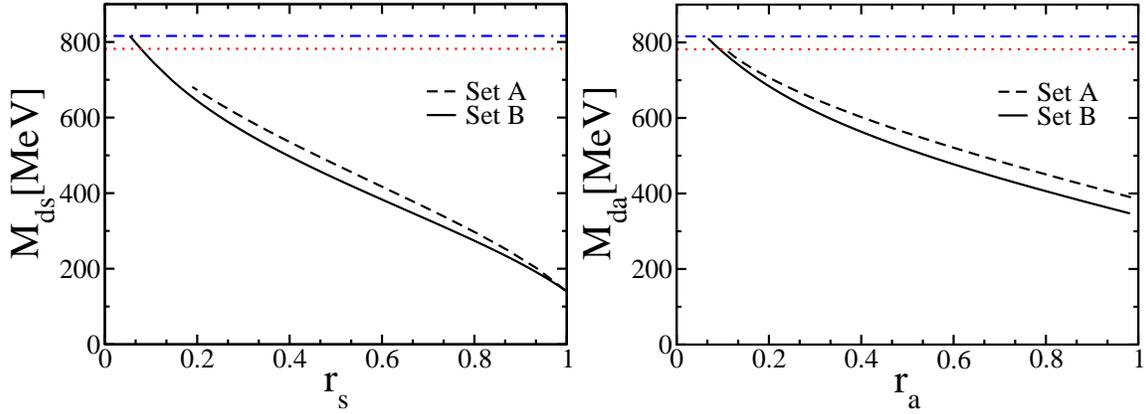

\vspace{1cm}
\includegraphics[height=.23\textheight]{plotsab.eps}
\includegraphics[height=.23\textheight]{plotab.eps}
\caption{The scalar and axial-vector diquark mass 
as a function of $r_{s}$ and $r_{a}$, respectively, for both parameter
sets. The dotted and the dash-dotted lines denote the quark-quark
pseudo-threshold for set A and B, respectively. \label{fig:samass}}
\end{figure}
\begin{figure}
     \centerline{\includegraphics[clip,width=7 cm] {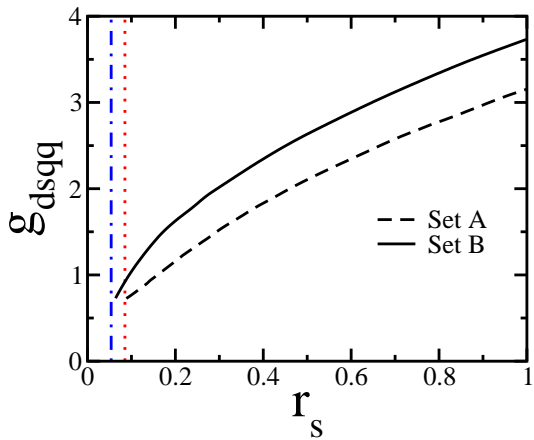}}
\caption{The scalar diquark-quark-quark coupling as a function of $r_{s}$.
The dotted and dash-dotted lines indicate the quark-quark pseudo-threshold for set A and B, respectively.\label{fig:gsdqq}}
\end{figure}
\begin{figure}
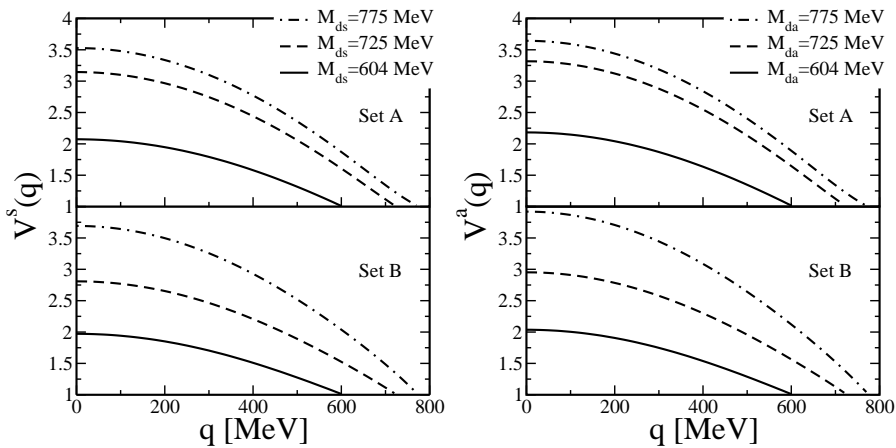

\begin{tabular}{cc}
\includegraphics[clip,height=.25\textheight]{plotoffshell-as.eps}&
\includegraphics[clip,height=.25\textheight]{plotoffshell-a.eps}
\end{tabular}
\caption{The ratio of the on-shell approximation compared to the 
exact diquark $T$-matrix for the various scalar and axial diquark
masses.\label{fig:offshellT}}
\end{figure}
\subsection{Diquark Solution}
The loop integrations in Eqs.~(\ref{l1},\ref{lon}) are very similar to
that appeared in mesonic sector Eq.~(\ref{pi2}). Therefore, we can employ the same
method to evaluate these loop integrations.

We use the parameter sets determined in the mesonic sector shown in
table \ref{tab:modpar}. Our numerical computation is valid below the
first quark-quark pseudo-threshold. Note that the longitudinal
polarizability $J^{L}_{a}(q)$ defined in Eq.~(\ref{lon}) does not
vanish here. This term will be neglected in our one-pole approximation
since it does not produce any poles in the $T$-matrix, and so makes a
very small contribution compared to the transverse piece (see
Fig.~\ref{fig:avT}). The longitudinal polarizability is not important
in the local NJL model as well \cite{w,njln3}. We find that for a wide
range of $r_{s}$ and $r_{a}$, for all parameter sets, a bound scalar
and axial-vector diquark exist (the results for additional sets can be
found in
\cite{me}). This is in contrast to the normal NJL model where a bound
axial-vector diquark exists only for very strong couplings
\cite{njln3}. The diquark masses for various values of $r_{s}$ and
$r_{a}$ are plotted in Fig.~\ref{fig:samass}. As already pointed out,
the scalar diquark mass is equal to the pion mass at $r_{s}=1$. It is
obvious from Fig.~\ref{fig:samass} that for $r_{s}=r_{a}$ the axial-vector diquark is heavier than the
scalar diquark, and consequently is rather loosely bound. For very
small $r_{s}$ and $r_{a}$, one finds
no bound state in either diquark channels.
In Fig.~\ref{fig:gsdqq}, we show the scalar diquark-quark-quark coupling
defined in Eq.~(\ref{co}) with respect to various scalar diquark couplings.

One should note that the nucleon bound state in the diquark-quark
picture does not require asymptotic-diquark states since the diquark
state is merely an intermediate device which simplifies the three-body
problem. Nevertheless, evidence for correlated diquark states in
baryons is found in deep-inelastic lepton scatterings and in hyperon weak decays \cite{lep2}. 
At the same time, diquarks appear as bound 
states in many phenomenological models, and are seen in lattice calculations
\cite{lat-diq1,lat-diq2}. In contrast to our perception of QCD colour
confinement, the corresponding spectral functions for these supposedly
confined objects in the colour anti-triplet channel are very similar to
mesonic ones \cite{lat-diq2}.

Next we study the off-shell behaviour of the diquark $T$-matrix. In
Fig.~\ref{fig:offshellT} we show the discrepancy between the exact
$T$-matrix and the on-shell approximation $V^{s,a}(q)$. At the pole we
have by definition that $V^{s,a}(q)|_{q^{2}=M^{2}_{s,a}}=1$.  We see
elsewhere that the off-shell contribution is very important due to the
non-locality of our model. We find that the bigger the diquark mass
is, the bigger the off-shell contribution.  The off-shell behaviour of
the scalar and the axial-vector channel for both parameter sets $A$
and $B$ are rather similar.

\section{Three-body sector\label{sec:TB}}
In order to make the three-body problem tractable, we discard any
three-particle irreducible graphs.  The relativistic Faddeev equation can be then
written as an effective two-body BS equation for a quark and a
diquark due to the locality of the form factor in momentum space (see
Eq.~(\ref{forfa})) and accordingly the separability of the two-body
interaction in momentum-space. We adopt the formulation developed by the
T\"ubingen group \cite{o1,o2,o3} to solve the resulting BS
equation. In the following we work in momentum space with Euclidean
metric. The BS wave function for the octet baryons can be presented in
terms of scalar and axialvector diquarks correlations,
\begin{equation}
\psi (p,P) u (P,s) =\left(\begin{array}{c}\psi^5 (p,P) \\ \psi^\mu(p,P) \end{array}\right)u(P,s), \label{bsn}
\end{equation}
where $u(P,s)$ is a basis of positive-energy Dirac spinors of spin $s$
in the rest frame. The parameters $p=(1-\eta)p_{i}-\eta(p_{j}+p_{k})$
and $P=p_{i}+p_{j}+p_{k}$ are the relative and total momenta in the
quark-diquark pair, respectively. The Mandelstam parameter $\eta$
describes how the total momentum of the nucleon $P$ is distributed
between quark and diquark.

One may alternatively define the vertex function associated with
$\psi (p,P)$ by amputating the external quark and diquark propagators (the legs) from the wave function;
\begin{equation}
\phi (p,P) =S^{-1}(p_{q})\tilde{D}^{-1}(p_{d})\left(\begin{array}{c}\psi^5 (p,P) \\ \psi^\nu(p,P) \end{array}\right), \label{psi}
\end{equation}
with 
\begin{equation}
\tilde{D}^{-1}(p_{d})=\left(\begin{array}{cc}D^{-1}(p_{d})&0\\0&(D^{\mu\nu}(p_{d}))^{-1},\end{array}\right) \label{di-pp}
\end{equation}
where $D(p), D^{\mu\nu}(p)$ and $S(p)$ are Euclidean versions of the
diquark and quark propagators which are obtained by the standard
transcription rules from the expressions in Minkowski space,
Eqs.~(\ref{n5},\ref{aloop}) and Eq.~(\ref{n3}), respectively. The
spectator quark momentum $p_{q}$  and the diquark momentum
$p_{d}$ are given by
\begin{eqnarray}
p_{q}&=&\eta P+p, \label{pk}\\
p_{d}&=&(1-\eta)P-p,\label{pkd}\
\end{eqnarray}
with similar expressions for $k_{q,d}$, where we replace $p$ by $k$
on the right-hand side. The Mandelstam
parameter $\eta$, parametrises different definitions of the relative momentum within the
quark-diquark system. In the ladder approximation, the coupled
system of BS equations for octet baryon wave functions and their
vertex functions takes the compact form,
\begin{figure}
       \centerline{\includegraphics[clip,width=11 cm]
                                   {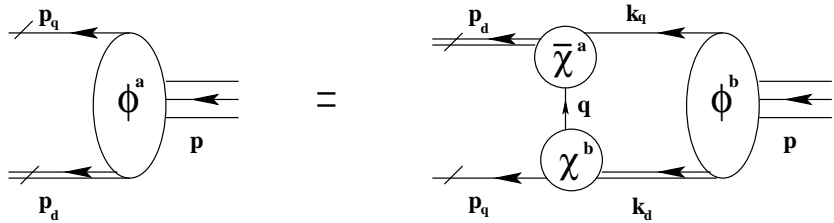}}
\caption{The coupled Bethe-Salpeter equation for the vertex function $\phi$.
\label{fig:BSPhi}}
\end{figure}
\begin{equation}
\phi (p,P)=\int \frac{d^{4}k}{(2\pi)^{4}} K^{BS}(p,k;P) \psi(k,P),\label{n7}\\
\end{equation}
where $K^{BS}(p,k;P)$ denotes the kernel of the nucleon BS equation
representing the exchange quark within the diquark with the spectator quark (see Fig.~\ref{fig:BSPhi}), and in the colour
singlet and isospin $\frac{1}{2}$ channel we find (see Ref.~\cite{njln3})
\begin{eqnarray}
K^{BS}(p,k;P)&=&-3\left(\begin{array}{cc}\chi^{5}(p_{1},k_{d})S^{T}(q)\bar{\chi}^{5}(p_{2},p_{d})&-\sqrt{3}\chi^{\alpha}(p_{1},k_{d})S^{T}(q)\bar{\chi}^{5}(p_{2},p_{d})\\-\sqrt{3}\chi^{5}(p_{1},k_{d})S^{T}(q)\bar{\chi}^{\mu}(p_{2},p_{d})&
-\chi^{\alpha}(p_{1},k_{d})S^{T}(q)\bar{\chi}^{\mu}(p_{2},p_{d})\end{array}\right),\label{n9}
\end{eqnarray}
where $\chi$ and $\chi^{\mu}$
(and their adjoint $\bar{\chi}$ and $\bar{\chi}^{\mu}$) stand for the
Dirac  structures of the scalar and the axial-vector diquark-quark-quark
vertices and can be read off  immediately from Eqs.~(\ref{tm1}, \ref{n5})
and Eqs.~(\ref{tm2}, \ref{aloop}), respectively. Therefore we have 
\begin{eqnarray}
\chi^{5}(p_{1},k_{d})&=&g_{dsqq}(\gamma^{5}C)\sqrt{2 V^{s}(k_{d})}f(p_{1}+(1-\sigma)k_{d})f(-p_{1}+\sigma k_{d}),\nonumber\\
\chi^{\mu}(p_{1},k_{d})&=&g_{daqq}(\gamma^{\mu}C)\sqrt{2 V^{a}(k_{d})}f(p_{1}+(1-\sigma)k_{d})f(-p_{1}+\sigma k_{d}),\label{d-v}\
\end{eqnarray}
where $\sigma$ is the Mandelstam parameter parameterising different
definitions of the relative momentum within the quark-quark system. We
have used an improved on-shell approximation for the contribution of
diquark $T$-matrix occurring in the Faddeev equations.  Instead of the
exact diquark $T$-matrices we use the on-shell approximation with a
correction of their off-shell contribution through $V^{s,a}(p)$.  What
is neglected is then the contribution to the $T$-matrix beyond the
pseudo-threshold.  We will see this approximation is sufficient to
obtain a three-body bound state. A similar approximation has been
already employed in the normal NJL model in the nucleon sector, see
for examples Refs.~\cite{cut-b,under2}. One should note that here we
do not have continuum states like the normal NJL model. However, there
exist many complex poles beyond the pseudo-threshold which may be
ignored, provided that they lie well above the energies of interest
and the cutoff. For the parameter sets considered here, the next set
of poles would result in another pseudo-threshold at energies of $1.3$
GeV and $3$ GeV for sets A and B, respectively. The model is not
intended to be reliable at such momenta. On the other hand, as we will
see in the next section, in practice, one may escape these poles far
enough away by taking advantage of the above Mandelstam
parametrization of the momenta.

The relative momentum of quarks in the diquark vertices $\chi$ and
$\chi^{\mu}$ are defined as $ p_{1}=p+k/2-(1-3\eta)P/2$ and
$p_{2}=-k-p/2+(1-3\eta)P/2$, respectively. The momentum $k_{d}$ of the
incoming diquark and the momentum $p_{d}$ of the outgoing diquark are
defined in Eq.~(\ref{pkd}) (see Fig.~\ref{fig:BSPhi}). The momentum of
the exchanged quark is fixed by momentum conservation at
$q=-p-k+(1-2\eta)P$.  In the expressions for the momenta we have
introduced two independent Mandelstam parameters $\eta, \sigma$, which
can take any value in $[0,1]$.  Observables should not depend on these
parameters if the formulation is Lorentz covariant. This means that
for every BS solution $\psi(p,P;\eta_{1},\sigma_{1})$ there exists an
equivalent family of solutions. This provides a stringent check on
calculations; see the next section for details.

It is interesting to note that the non-locality of the diquark-quark-quark
vertices naturally provides a regularisation of the
ultraviolet divergence in the diquark-quark loop.

We now constrain the Faddeev amplitude to describe a state of
positive energy, positive parity and spin $s=1/2$. The  parity
condition can be immediately reduced to a
condition for the BS wave function:
\begin{equation}
\mathcal{P}\left(\begin{array}{c}\psi^{5}(p,P)\\ \psi^{\mu}(p,P)\end{array}\right)=\left(\begin{array}{c}\gamma^{4}\psi^{5}(\bar{p},\bar{P})\gamma^{4}\\ \gamma^{4}\Lambda^{\mu\nu}_{\mathcal{P}}\psi^{\nu}(\bar{p},\bar{P})\gamma^{4} \end{array}\right)=\left(\begin{array}{c}\psi^{5}(p,P)\\ -\psi^{\mu}(p,P)\end{array}\right), 
\end{equation}
where we define $\bar{p}=\Lambda_{\mathcal{P}}p$ and $
\bar{P}=\Lambda_{\mathcal{P}}P$, with
$\Lambda^{\mu\nu}_{\mathcal{P}}=\text{diag}(-1,-1,-1,1)$. In order to
ensure the positive energy condition, we project the BS wave function
with the positive-energy projector
\(\Lambda^{+}=(1+\hat{\slashP})\), where the hat
denotes a unit four vector (in rest frame we have
$\hat{P}=P/iM$). Now we expand the BS wave function $\psi(p,P)$ in
Dirac space $\Gamma\in\{\textbf{1}, \gamma_{5},\gamma^{\mu},
\gamma_{5}\gamma^{\mu},\sigma^{\mu\nu}\}$. The above-mentioned
conditions reduce the number of independent component from sixteen to
eight, two for the scalar diquark channel, $S_{i},
(i=1,2)$ and six for the axial-diquark channel, $A_{i},(i=1,...6)$.
The most general form of the BS wave function is given by
\begin{eqnarray}
\psi^{5}(p,P)&=&\left(S_{1}-i\hat{\slashp}_{T}S_{2}\right)\Lambda^{+},\nonumber\\
\psi^{\mu}(p,P)&=&\Big(i\hat{P}^{\mu}\hat{\slashp}_{T}A_{1}+\hat{P}^{\mu}A_{2}-\hat{p}^{\mu}_{T}\hat{\slashp}_{T}A_{3}+
i\hat{p}^{\mu}_{T}A_{4}+\left(\hat{p}^{\mu}_{T}\hat{\slashp}_{T}-\gamma^{\mu}_{T}\right)A_{5}\nonumber\\
&&-(i\gamma^{\mu}_{T}\hat{\slashp}_{T}+i\hat{p}^{\mu}_{T})A_{6}\Big)\gamma_{5}\Lambda^{+}.
\label{main}\
\end{eqnarray}
Here we write
\(\gamma^{\mu}_{T}=\gamma^{\mu}-\hat{\slashP}\hat{P}^{\mu}\). 
The subscript $T$ denotes the component of a four-vector transverse to
the nucleon momentum, $p_{T}=p-\hat{P}(p.\hat{P})$. In the same way,
one can expand the vertex function $\phi$ in Dirac space, and since
the same constraints apply to the vertex function, we obtain an
expansion similar to Eq.~(\ref{main}), with new unknown coefficients
$\mathbb{S}_{i}$ and $\mathbb{A}_{i}$. The unknown functions $S_{i}
(\mathbb{S}_{i})$ and $A_{i} (\mathbb{A}_{i})$ depend on the two
scalars which can be built from the nucleon momentum $P$ and relative
momentum $p$, $z=\hat{P}.\hat{p}=\cos\omega$ (the cosine of the
four-dimensional azimuthal angle of $p^{\mu}$) and $p^{2}$. Of course,
they depend on $P^{2}$ as well, but this dependence becomes trivial
in the nucleon rest frame.

In the nucleon rest frame, one can rewrite the Faddeev amplitude in
terms of tri-spinors each possessing definite orbital angular momentum
and spin \cite{o1}. It turns out that these tri-spinors can be written
as linear combinations of the eight components defined in
Eq.~(\ref{main}).  Thus from knowledge of $S_{i}$ and $A_{i}$, a full
partial wave decomposition can be immediately obtained
\cite{o1}. Note that the off-shell contribution $V^{s,a}(q)$ is a function of the scalar
$q^{2}$.  Moreover, the form factor in our model Lagrangian is also
scalar, hence the total momentum dependent part of the
diquark-quark-quark vertices are scalar functions and carry no orbital
angular momentum, i.e. $L^{2}\chi^{5,\mu}(q)=0$. Therefore, the
partial wave decomposition obtained in Ref.~\cite{o1} for pointlike
diquarks can be used here. Notice that no such partial wave
decomposition can be found for the BS vertex function $\phi^{5,\mu}$
since the axial-vector diquark propagator mixes the space component of
the vertex function and time component of the axial-vector diquark.

\subsection{Numerical method for the coupled BS equations}
To solve the BS equations we use the algorithm introduced by Oettel
{\em et al} \cite{o4} . The efficiency of this algorithm has already been
reported in several publications, see for example
Refs.~\cite{o1,o2,o3}.  We will focus here only on the key ingredients
of this method. The momentum dependence of quark mass in
our model increases the complexity of the computation significantly.
As usual, we work in the rest frame of the nucleon $P=(0,iM_{N})$. In
this frame we are free to chose the spatial part of the relative
momentum $p$ parallel to the third axis. Thus the momenta
$p$ and $k$ are given by
\begin{eqnarray} 
p^{\mu}&=&|p|(0,0,\sqrt{1-z^{2}},z), \nonumber\\ 
k^{\mu}&=&|k|(\sin\theta'\sin\phi'\sqrt{1-z'^{2}},\sin\theta'\cos\phi'\sqrt{1-z'^{2}},\cos\theta\sqrt{1-z'^{2}},z'),\
\end{eqnarray}
where we write $z=\cos\omega$ and $z'=\cos\omega'$. 
The wave function Eq.~(\ref{main}) consists of $2\times2$ blocks in Dirac space can be simplified to
\begin{eqnarray}
\psi^{5}(p,P)&=&\left(\begin{array}{cc}
S_{1}(p^{2}, z) & 0 \\
\sigma_{3}\sqrt{1-z^{2}}S_{2}(p^{2},z) &0
\end{array}\right),
\hspace{1.2cm}
\psi^{4}(p,P)=\left(\begin{array}{cc}
\sigma_{3}\sqrt{1-z^{2}}A_{1}(p^{2},z) & 0 \\
A_{2}(p^{2},z) & 0\end{array}\right),   \nonumber\\
\psi^{3}(p,P)&=&\left(\begin{array}{cc}
i\sigma_{3}A_{3}(p^{2},z) & 0 \\
i\sqrt{1-z^{2}}A_{4}(p^{2},z) & 0\end{array}\right),
\hspace{1.5cm}
\psi^{2}(p,P)=\left(\begin{array}{cc}
i\sigma_{2}A_{5}(p^{2},z) & 0 \\
-\sigma_{1}\sqrt{1-z^{2}}A_{6}(p^{2},z) & 0\end{array}\right),\nonumber\\
\psi^{1}(p,P)&=&\left(\begin{array}{cc}
i\sigma_{1}A_{5}(p^{2},z) & 0 \\
\sigma_{2}\sqrt{1-z^{2}}A_{6}(p^{2},z) & 0\end{array}\right).\label{rest-bs}\
\end{eqnarray}
The great advantage of this representation is that the scalar and the
axial-vector components are decoupled. Therefore the BS equation
decomposes into two sets of coupled equations, two for the scalar
diquark channel and six for the axial diquark
channel.  We expand the vertex (wave) functions in terms of Chebyshev
polynomials of the first kind, which are closely related to the
expansion into hyperspherical harmonics. This decomposition turns
out to be very efficient for such problems \cite{o1,o2,o3,o4}. Explicitly, 
\begin{eqnarray}
F^{\psi}_{i}(p^{2},z)&=&\sum^{n_{\text{max}}}_{n=0}i^{n}F^{\psi(n)}_{i}(p^{2})T_{n}(z),\label{che}\nonumber\\
F^{\phi}_{i}(p^{2},z)&=&\sum^{m_{\text{max}}}_{m=0}i^{n}F^{\phi(m)}_{i}(p^{2})T_{m}(z),\
\end{eqnarray}
where $T_{n}(z)$ is the Chebyshev polynomial of the first kind. We use
a generic notation where the functions $F^{\psi}_{i}$(and $F^{\phi}_{i}$)
stand for any of the functions $S_{i},~ A_{i}$ (and
$\mathbb{S}_{i},~\mathbb{A}_{i}$),
\begin{eqnarray}
&& S_{1,2}\to F^{\psi}_{1,2}, \hspace{2cm}  A_{1...6}\to F^{\psi}_{3...8}, \nonumber\\
&& \mathbb{S}_{1,2}\to F^{\phi}_{1,2}, \hspace{2cm}  \mathbb{A}_{1...6}\to F^{\phi}_{3...8} .
\end{eqnarray}
We truncate the Chebyshev
expansions involved in $F^{\psi}_{i}$ and $F^{\phi}_{i}$ at
different orders $n_{\text{max}}$ and $m_{\text{max}}$, respectively. We also expand
the quark and diquark propagators into Chebyshev polynomials. In this
way one can separate the $\hat P\cdot \hat p$ and $\hat P \cdot \hat
k$ dependence in Eqs.~(\ref{psi},\ref{n7}). Using the
orthogonality relation between the Chebyshev polynomials, one can
reduce the four dimensional integral equation into a system of coupled
one-dimensional equations. Therefore one can rewrite Eqs.~(\ref{psi},\ref{n7}) in the matrix form
\begin{eqnarray}
F^{\psi(n)}_{i}(p^{2})&=&\sum_{j=1}^{8}\sum_{m=0}^{m_{\text{max}}}g_{ij}^{nm}(p^{2})F^{\phi(m)}_{j}(p^{2}),\nonumber\\
F^{\phi(m)}_{i}(p^{2})&=&\sum_{j=1}^{8}\sum_{n=0}^{n_{\text{max}}}\int^{\infty}_{0}d|k| |k|^{3} H_{ij}^{mn}(k^{2},p^{2})F^{\psi(n)}_{j}(k^{2}).\label{bs-matrix}\
\end{eqnarray}
Here $g_{ij}^{nm}$ and $H_{ij}^{mn}$ are the matrix elements of the
propagator and the quark-exchange matrices, respectively. The indices
$n, m$ refer to the Chebyshev moments and $i, j$ denote the individual channels.
To solve  Eq.~(\ref{bs-matrix}), we first rewrite it in the 
form of linear eigenvalue problem. Schematically
\begin{equation}
\lambda(P^{2})\varphi=K(P^{2})\varphi,
\end{equation}   
with the constraint that $\lambda(P^{2})=1$ at $P^{2}=-M^{2}_{N}$. This
can be used to determine the nucleon mass $M_{N}$ iteratively.
\begin{figure}[!ht]
   \centerline{\includegraphics[clip,width=7cm]
                                   {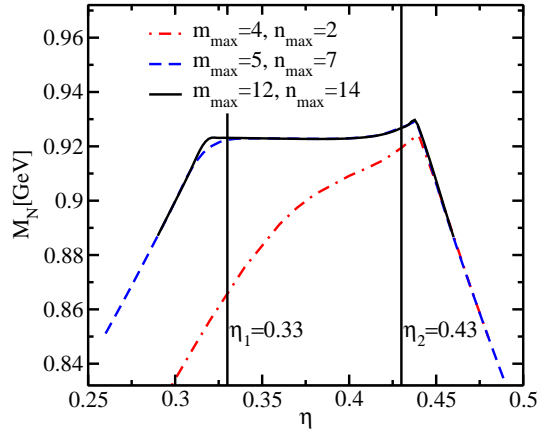}}
\caption{The dependence of the nucleon mass on the Mandelstam parameter $\eta$
for three choices of the cut-off on the Chebyshev expansion. Here we use set B,
with $M_{ds}=725 $ MeV and $M_{da}=630$ MeV. The vertical lines at $\eta_1$
and $\eta_2$ denote the position of the
singularities defined in Eq.~(\ref{sin}).\label{fig:Mandelstam}}
\end{figure}

As already pointed out, the BS solution should be independent of the
Mandelstam parameters $\eta,\sigma$. As can be seen in Fig.~\ref{fig:Mandelstam}, there is indeed a large plateau for the $\eta$ dependence if we use
a high cut-off on the Chebyshev moments. The limitations on the
size of this area of stability can be understood by considering where
the calculation contains singularities due to quark and diquark poles,
\begin{eqnarray}
\eta&\in&\left[1-\frac{M_{ds}}{M_{N}}, \frac{\mqr }{M_{N}}\right], \hspace{2cm}\text{if} \hspace{1cm} M_{ds}<M_{da},\nonumber\\
\eta&\in&\left[1-\frac{M_{da}}{M_{N}}, \frac{\mqr }{M_{N}}\right],  \hspace{2cm}\text{if} \hspace{1cm} M_{da}<M_{ds}.\label{sin}
\end{eqnarray}
A similar plateau has been found in other applications \cite{o1,o2,o3}. The singularities in the quark-exchange propagator
put another constraint on the acceptable range of $\eta$;
$\eta>\frac{1}{2}(1-\frac{\mqr }{M_{N}})$. No such constraint exists
for $\sigma$, which relates to the relative momentum between two
quarks. To simplify the algebra we take $\sigma=1/2$.

In what follows we use a momentum mesh of $60\times 60$ for $p,k$,
mapped in a non-linear way to a finite interval. In the non-singular
regime of Mandelstam parameter $\eta$ Eq.~(\ref{sin}), the Faddeev
solution is almost independent of the upper limit on the Chebyshev
expansion, and for $m_{\text{max}}=10, n_{\text{max}}=12$, see the
Fig.~\ref{fig:Mandelstam}, this seems to be satisfied. This limit is
somewhat  higher than the reported values for simple models \cite{o1,o2,o3,o4}.
\begin{figure}
 \centerline{\includegraphics[clip,width=7 cm]{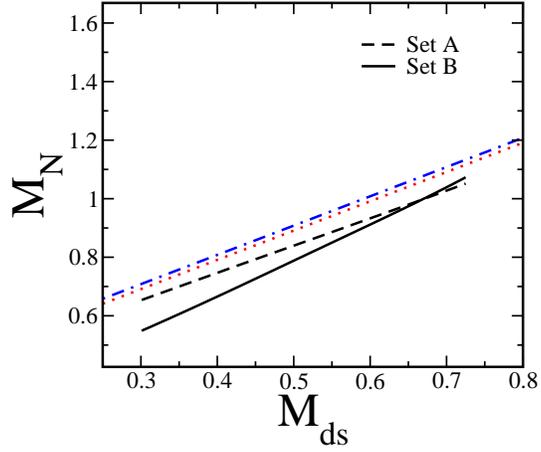}}
\caption{The nucleon mass without inclusion of the axial diquark channel. 
The dotted lines indicate the diquark-quark threshold. All values are
given in GeV.\label{fig:Nmass_scal}}
\end{figure}
\subsection{Nucleon Solution}
In order to understand the role of the axial diquark in nucleon
solution, we first consider the choice $r_{a}=0$. For this case we
find that the non-confining set A can not generate a three-body bound
state without the inclusion of the off-shell contribution. For the
confining set B one also has to enhance the diquark-quark-quark
coupling $g_{dsqq}$ by a factor of about $1.73$ over the value defined
in Eq.~(\ref{co}) (as we will show, this extra factor is not necessary
when the axial-vector diquark is included).  The situation is even
more severe in the on-shell treatment of the local NJL model, since
one needs to include the quark-quark continuum contribution in order
to find a three-body bound state when the axial-vector diquark channel
is not taken into account \cite{njln1}.
\begin{figure}[!t]
       \centerline{\includegraphics[clip,width=7 cm]
                                  {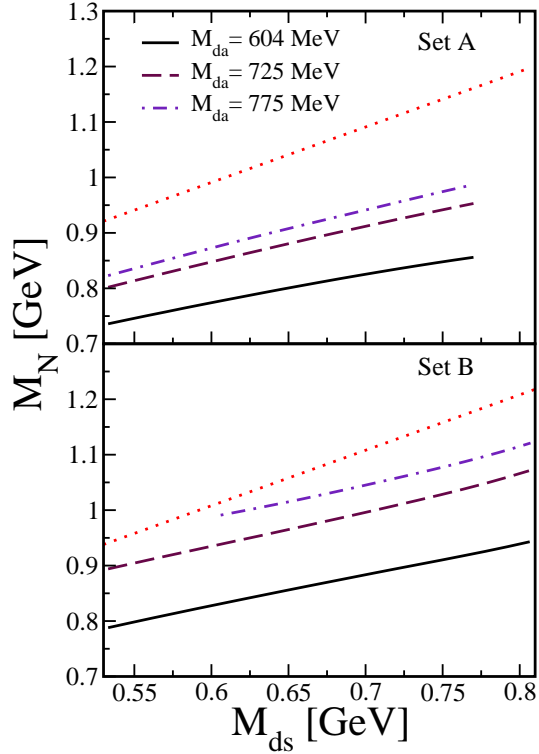}}
\caption{\label{fig:Nmass_full2} 
The nucleon mass as a function of the scalar 
diquark mass for various axial vector diquark masses for both parameter sets. The scalar diquark-quark threshold are shown by the dotted lines.}
\end{figure}
\begin{figure}[!t]
       \centerline{\includegraphics[clip,width=7 cm]
                                   {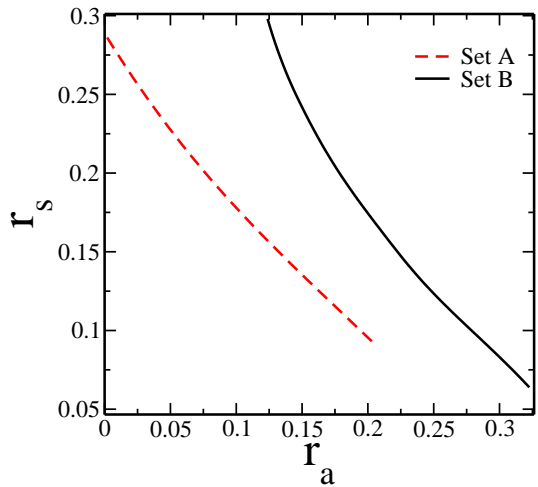}}
\caption{Range of parameters $(r_{s},r_{a})$ where we find a nucleon mass of 
$940$ MeV.\label{fig:Nmass_phys}}
\end{figure}

As can be seen from Fig.~\ref{fig:samass} a decrease in $r_{s}$ leads
to a larger diquark mass, and an increase in the off-shell
contribution to the quark-quark $T$-matrix (see
Fig.~\ref{fig:offshellT}). This off-shell correction is crucial
for forming a bound nucleon. If the off-shell contribution
$V^{s}(q)$ is omitted, a bound nucleon can not be found.
 
\begin{table}
\caption{Diquark masses and coupling of diquarks to quarks
obtained for $M_{N}=940$ MeV. All masses are given in MeV.
 $E_{ds}(E_{da})$ denote the binding energy of diquarks in the
 nucleon, $E^{ds}_{N}(E^{da}_{N})$ denote the binding energy of the
 nucleon measured from scalar (axial) diquark mass.\label{tab:mdiquark}}
\begin{ruledtabular}
\begin{tabular}{lllllll}
\multicolumn{1}{c}{}& \multicolumn{3}{c}{Set A}& \multicolumn{3}{c}{Set B}  \\
&Set A1 & Set A2 & Set A3 & Set B1 &Set B2 &Set B3\\
\hline
$M_{ds}$& 775     & 748   &698       & 802        &705  &609        \\
$g_{dsqq}$&0.74    &0.83   &1.04       & 0.73       &  1.31    &1.79       \\
$r_{s}$ &0.09      &0.12  &0.17       & 0.06         &  0.14   &0.24        \\
$E_{ds}$& 7 &34  &84       & 14   & 111   &207          \\
$E^{ds}_{N}$&226  &199 &149        & 270   & 173   &77           \\
$M_{da}$&705      &725 &775       & 604       & 660  &725       \\
$g_{daqq}$&1.08     &0.98  &0.79         & 1.99      & 1.67   &1.28      \\
$r_{a}$&0.20        &0.17  &0.11           & 0.32        & 0.23   &0.15     \\
$E_{da}$&77  &57  &7     & 212  & 156   &91       \\
$E^{da}_{N}$&156  &176 &226          & 72  & 123 &193       \\
$p_{\bot}^{\text{RMS}}$&194.88 &181.51 &163.70 &283.99& 232.86 &  209.95 \\
\end{tabular}
\end{ruledtabular}
\end{table}

The nucleon result is shown in Fig.~\ref{fig:Nmass_scal}. We also show
a fictitious diquark-quark threshold defined as $M_{ds}+\mqr$. The
nucleon mass can be seen to depend roughly linearly on the scalar
diquark mass. A similar behaviour is also seen in the local NJL model
\cite{njln2}. Increasing the diquark mass (or decreasing $r_{s}$)
increases the nucleon mass, i.e. the scalar diquark channel is
attractive.  In order to obtain a nucleon mass of $940 $ MeV, we need
diquark mases of $608$ MeV and $623$ MeV for set A and B,
respectively. The corresponding nucleon binding energy measured from
the diquark-quark threshold are $ 56$ MeV and $ 91$ MeV for set A and
B, respectively, compared to the binding of the diquarks (relative to
the quark-quark pseudo threshold) of about $174$ and $193$ MeV for set A and
B, respectively. Such diquark clustering within the nucleon is also
observed in the local NJL model
\cite{njln2}, and is qualitatively in agreement with an instanton
model \cite{i-b} and lattice simulations
\cite{lat-bin}.

Next we investigate the effect of the axial-vector diquark channel on
nucleon solution. We find that the axial-vector diquark channel
contributes considerably to the nucleon mass and takes away the need for the artificial enhancement of the
coupling strength for set B. In Fig.~\ref{fig:Nmass_full2} we show the nucleon mass as a function of
the scalar and axial-vector diquark mass.  As in the scalar diquark
channel, we define the axial-vector diquark-quark threshold as
$M_{da}+\mqr$. We see that as one increases the axial-vector diquark
(and scalar diquark) masses, the quark-quark interaction is weakened and
consequently the nucleon mass increases. Therefore the contribution
of the axial-vector channel to the nucleon mass is also attractive.
\begin{figure}
\centerline{\includegraphics[width=11 cm]{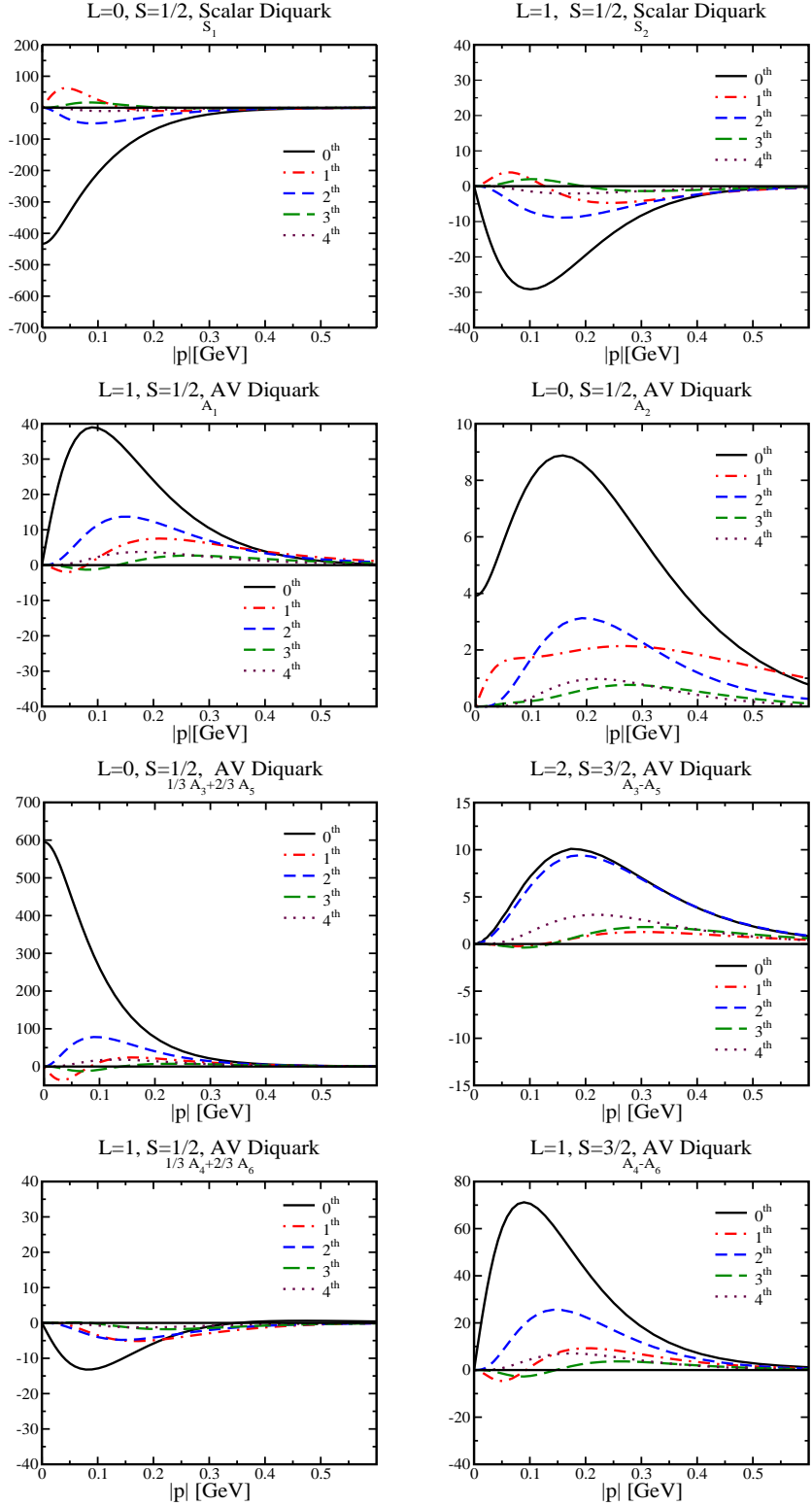}}
\caption{Chebyshev moments (labelled by $n$) of the scalar and axial-vector (AV) diquark amplitudes of the nucleon BS wave function given by Set A1.\label{fig:ChebA}}
\end{figure}

\begin{figure}
\centerline{\includegraphics[width=11 cm]{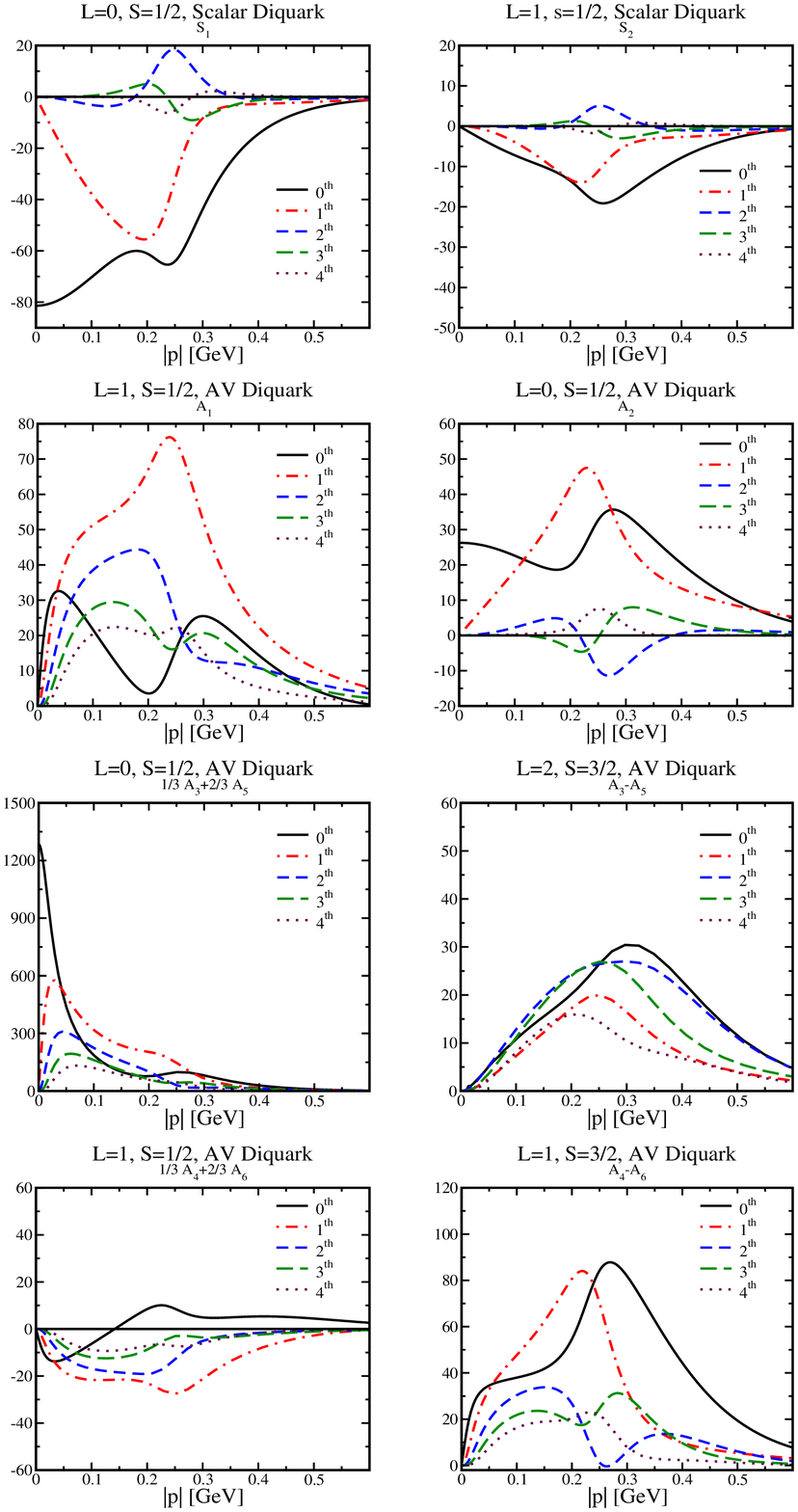}}
\caption{Chebyshev moments (labelled by the order $n$) of scalar and 
axial-vector (AV) diquark amplitudes for the nucleon BS wave function
obtained for parameter set B1.\label{fig:ChebB}}
\end{figure}

In Fig.~\ref{fig:Nmass_phys} we plot the parameter space of the
interaction Lagrangian with variable $r_{s}$ and $r_{a}$ which leads
to the nucleon mass $M_{N}=940$ MeV. The trend of this plot for the
non-confining set A is very similar to the one obtained in the local
NJL model \cite{njln3}.

If the scalar diquark interaction $r_{s}$ is less than $0.14$, we need
the axial-vector interaction to be stronger than the scalar diquark
channel $r_{a}>r_{s}$ in order to get the experimental value of
nucleon mass. For set B, as we approach $r_{a}=0$, the curve bends
upward, reflecting the fact that we have no bound state with only the
scalar diquark channel.

In Fig.~\ref{fig:Nmass_phys} we see for the confining set B that the
interaction is again shared between the scalar and the axial-vector
diquark and for small $r_{s}<0.19$ one needs a dominant axial-vector
diquark channel $r_{a}>r_{s}$. It is obvious that the axial-vector
diquark channel is much more important in the confining than the
non-confining phase of model. 
\begin{figure}
       \centerline{\includegraphics[clip,width=7 cm]
                                   {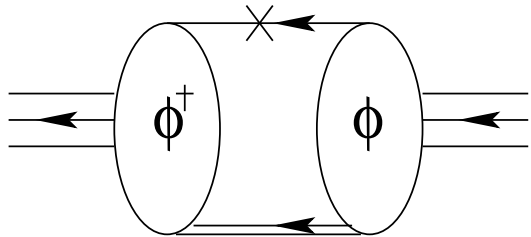}}
\caption{Diagram corresponding to the definition of density Eq.~(\ref{den-is}).
\label{fig:n2}}
\end{figure}

In order to study the implications of the quark confinement for the
description of the nucleon, we compare in Table \ref{tab:mdiquark} three
representative cases for both the non-confining and confining
parameter sets, which all give a nucleon mass of about $940$ MeV.  The
first three columns contain results for set $A$, and the last three
columns for the confining set $B$.

Given the definition of diquark-quark thresholds, in the presence of
both scalar and axial-vector diquark channels, the diquarks in the
nucleon can be found much more loosely bound, although one obtains a
very strongly bound nucleon solution near its experimental value, see
table \ref{tab:mdiquark}. Next we study the nucleon BS wave function
for the various sets given in Table \ref{tab:mdiquark}. The nucleon
wave and vertex function are not physical observables, but rather they
suggest how observables in this model will behave. In
Figs.~\ref{fig:ChebA}, \ref{fig:ChebB} we show the leading Chebyshev
moments of the scalar functions of the nucleon BS wave function for
various sets (A1 and B1). They describe the strengths of the
quark-diquark partial waves with $S$ as a total quark-diquark spin and
$L$ as a total orbital angular momentum. They are normalised to
$F_{1}^{\phi(0)}(p_{1})=1$, where $p_{1}$ is the first point of the
momentum mesh. Very similar plots are found for the other sets of
parameters given in table~\ref{tab:mdiquark}. It is seen that the
contribution of higher moments are considerably smaller than lower ones, indicating a
rapid convergence of the expansion in terms of
Chebyshev polynomials. In the confining case, Fig.~\ref{fig:ChebB},
there is a clear interference which is not present in the
non-confining one, Fig.~\ref{fig:ChebA}. Therefore in the confining case,
all wave function amplitudes are shifted to higher relative
four-momenta between the diquark and quark. 

\begin{figure}[!t]
\centerline{\includegraphics[width=14 cm]{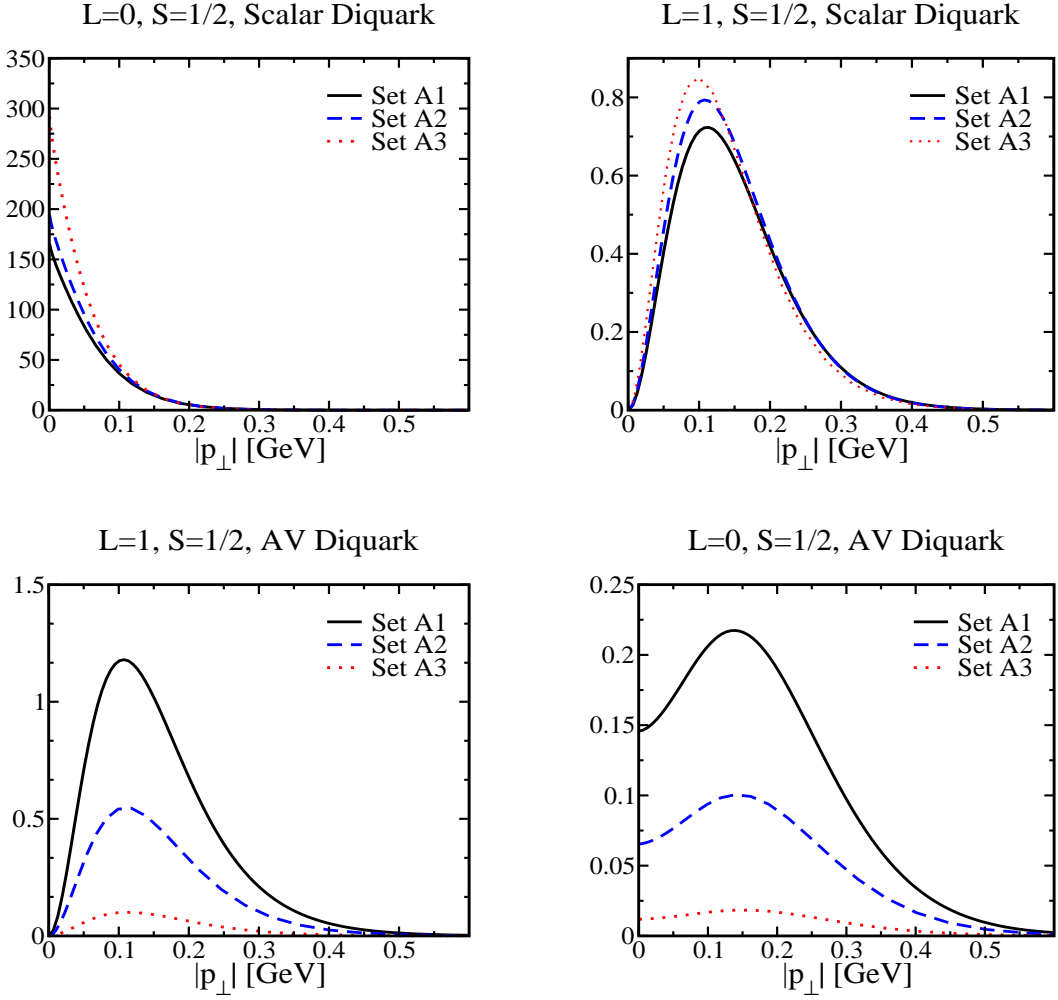}}
\caption{Shows the nucleon density ($M=940$ MeV, Set $A$) with  respect to relative
momentum between diquark and quark for different set of $A1, A2$ and $A3$.
\label{fig:bdensA}}
\end{figure}

\begin{figure}[!t]
\centerline{\includegraphics[width=14 cm]{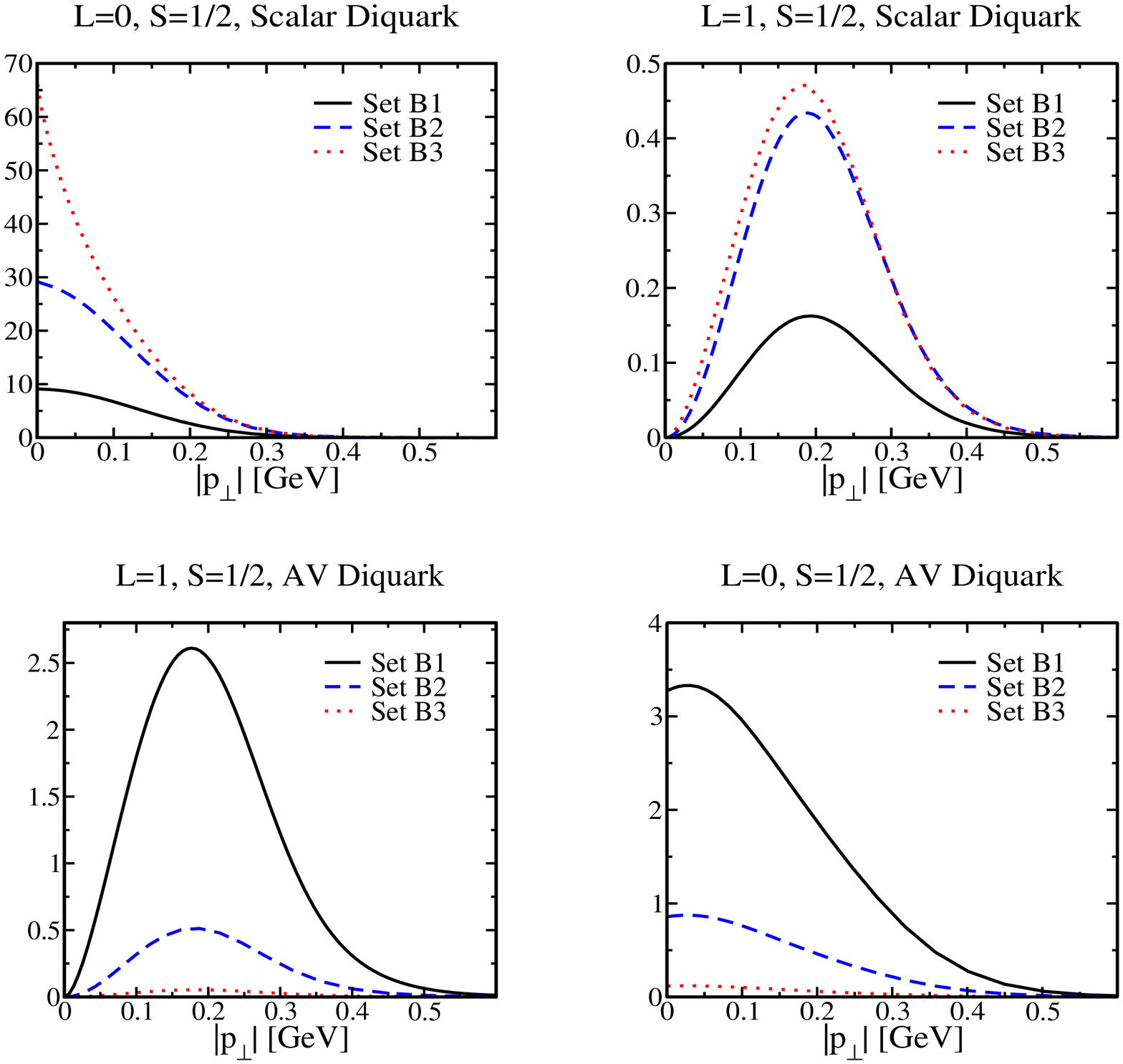}}
\caption{Shows the nucleon density ($M=940$ MeV, Set $B$) with respect to relative
momentum between diquark and quark for different set of $B1, B2$ and $B3$. \label{fig:bdensB}}
\end{figure}

In order to understand the effect of this interference, we construct 
a density function for the various channels in the nucleon rest frame. This density
is defined as
\begin{equation}
\rho(p_{\bot},P)=\int dp_{4}\psi^{\dag}(p_{\bot},p_{4},P)\tilde{D}^{-1}(p_{d})\psi(p_{\bot},p_{4},P),  \label{den-is}
\end{equation}
where $p_{\bot}$ stands for the space component of the relative
momentum $p$, and $\tilde{D}^{-1}(p_{d})$ is defined in
Eq.~(\ref{di-pp}). This definition corresponds to a very naive diagram
describing the quark density within the nucleon, see Fig.~\ref{fig:n2}. In the above definition
of the density function, we have integrated over the time component of
the relative momentum. In this way the density function becomes very
similar to its counterpart in Minkowski space. Although the above
definition of density is not unique, it does provide a useful measure
of the spatial extent of the wave function. Similar calculations have been done for the quark
condensate in Ref.~\cite{njln4-is}. The results are plotted in
Figs.~\ref{fig:bdensA} and \ref{fig:bdensB}.

It is noticeable that in all cases, the $s$-wave in scalar diquark channel is the dominant
contribution to the ground state. The relative importance of the
scalar and the axial diquark amplitude in the nucleon changes with the
strength of the diquark-quark couplings $g_{dsqq}(g_{daqq})$ and
accordingly with $r_{s}(r_{a})$. There are indications that in the confining sets,
the nucleon density extends to higher relative momentum between the
diquark and the quark. These imply a more compact nucleon in the
confining cases. In order to find a quantitative estimate of the
confinement effect in our model, we calculate $ p_{\bot}^{\text{RMS}}=(\langle p^{2}_{\bot}\rangle-\langle p_{\bot}
\rangle^{2})^{1/2}$; the results can be found in table \ref{tab:mdiquark}. 

We also see in the both confining and non-confining cases a decrease in
$p_{\bot}^{\text{RMS}}$ with weakening axial-vector diquark
interaction (and consequently increasing the scalar diquark
interaction strength). This can be associated with the important role of the axial-vector diquarks. If we compare $p_{\bot}^{\text{RMS}}$ for the
two sets $A2$ and $B2$, which have very similar interaction parameters
$r_{s}(r_{a})$, an increase about $25\%$ is found.

\section{Summary and Outlook\label{sec:conc}}
In this work we have investigated the two- and three-quark problems in
a non-local NJL model. We have truncated the diquark sector to the
scalar and the axial-vector channels. We have solved the relativistic
Faddeev equation for this model and have studied the behaviour of the
nucleon solutions with respect to various scalar and the axial-vector
interactions. We have studied the dependence of the baryon masses and waves
on the interaction parameters $r_{s}$ and $r_{a}$ (and on the scalar and the axial-vector diquark mases), which can describe
the pion and the nucleon simultaneously. In order to put more restrictions on these parameters, one would need to calculate the $\Delta$ mass as well. 

Although the model is quark confining, it is not diquark confining (at
least in the rainbow-ladder approximation). A bound diquark can be
found in both scalar and the axial-vector channel for a wide range of
couplings. We have found that the off-shell contribution to the
diquark $T$-matrix is crucial for the calculation of the nucleon:
without it the attraction in the diquark channels is too weak to form
a three-body bound state. We have also found that both the scalar and
the axial-vector contribute attractively to the nucleon mass. The role
of axial-vector channel is much more important in the confining phase
of model.  The nucleon in this model is strongly bound even though the
diquarks are rather loosely bound. The confining aspects of the model
are more obvious in three-body, rather than the two-body sector. We
decomposed the nucleon BS wave function in the nucleon rest
frame in terms of spin and orbital angular momentum eigenstates and
constructed the quark density function in the various channels. It was revealed that $s$-wave in scalar
diquark channel is the dominant contribution to the ground state.
By investigating the nucleon wave function we found that quark
confinement leads to a more compact nucleon. The size of nucleon is
reduced by about $25\%$ in the confining cases.

For both confining and non-confining cases, an increase in the scalar
diquark channel interaction $r_{s}$ leads to a lower nucleon mass, see
Fig.~\ref{fig:Nmass_full2}. However, the mass of the $\Delta$ should be
independent of $r_{s}$ since it does not contain scalar diquarks
\cite{njln3,o1,i-b}. In the standard NJL model the difference between the
nucleon and $\Delta$ masses is strongly dependent on the scalar diquark
interaction \cite{njln3}. In the current model where the axial-vector
diquark makes a larger contribution to the nucleon mass, a
detailed calculation of the $\Delta$ is needed to understand
the mechanism behind the $\Delta$-$N$ mass difference.  Note also that
neglecting $\pi N$-loops may lead to a quantitative overestimate of
the axial-vector diquark role in the nucleon
\cite{pion-n}.

In order to understand the implications of this model in baryonic
sector fully one should investigate other properties of the nucleons such as
the charge radii, magnetic moments and axial coupling. On the
other hand, the role of quark confinement in this model may be better
clarified by investigating quark and nuclear matter in this
model. Such problems can also be studied within the same Faddeev approach
\cite{w,njl-av}.

\section*{Acknowledgements}
One of the authors (AHR) would like to thank R. Plant for useful correspondence at an early stage of this work. 
The research of AHR was supported by a British Government ORS award and a UMIST grant. The work of NRW and MCB was supported by the UK EPSRC under grants GR/N15672 and GR/N15658. 

\end{document}